\documentclass[fleqn,twoside]{article}
\usepackage{espcrc2}

\usepackage{graphicx}
\usepackage[figuresright]{rotating}

\usepackage{amsmath}
\usepackage{amssymb}
\newcommand{\be}{\begin{equation}}
\newcommand{\ee}{\end{equation}}
\newcommand{\bea}{\begin{eqnarray}}
\newcommand{\eea}{\end{eqnarray}}

\title{Finite density QCD with a canonical approach}

\author{
Philippe de Forcrand\address[ETH]{Institute for Theoretical Physics, 
ETH Z\"{u}rich,  CH-8093 Z\"{u}rich, Switzerland}
\address{Physics Department, TH Unit, CERN, CH-1211 Geneva 23, Switzerland}
and
Slavo Kratochvila\addressmark[ETH]
}
       
\begin{document}

\begin{abstract}
We present a canonical method where the properties of QCD are directly
obtained as a function of the baryon density $\rho$, rather than the
chemical potential $\mu$. We apply this method to the determination of
the phase diagram of four-flavor QCD. For a pion mass $m_\pi \sim 350$ MeV, 
the first-order transition between the hadronic and the plasma phase gives 
rise to a co-existence region in the $T$-$\rho$ plane, which we study in detail,
including the associated interface tension.
We obtain accurate results for systems containing up to 30 baryons
and quark chemical potentials $\mu$ up to $2 T$.
Our $T$-$\mu$ phase diagram agrees with the literature
when $\frac{\mu}{T}\lesssim 1$.
At larger chemical potential, we observe a ``bending down'' of
the phase boundary. We compare the free energy in the confined and deconfined
phase with predictions from a hadron resonance gas and from a free massless
quark gas respectively.

\end{abstract}

\maketitle

\section{Introduction}
                                                                                
\begin{figure}[!htb]
\begin{center}
\includegraphics[height=6.00cm,angle=-90]{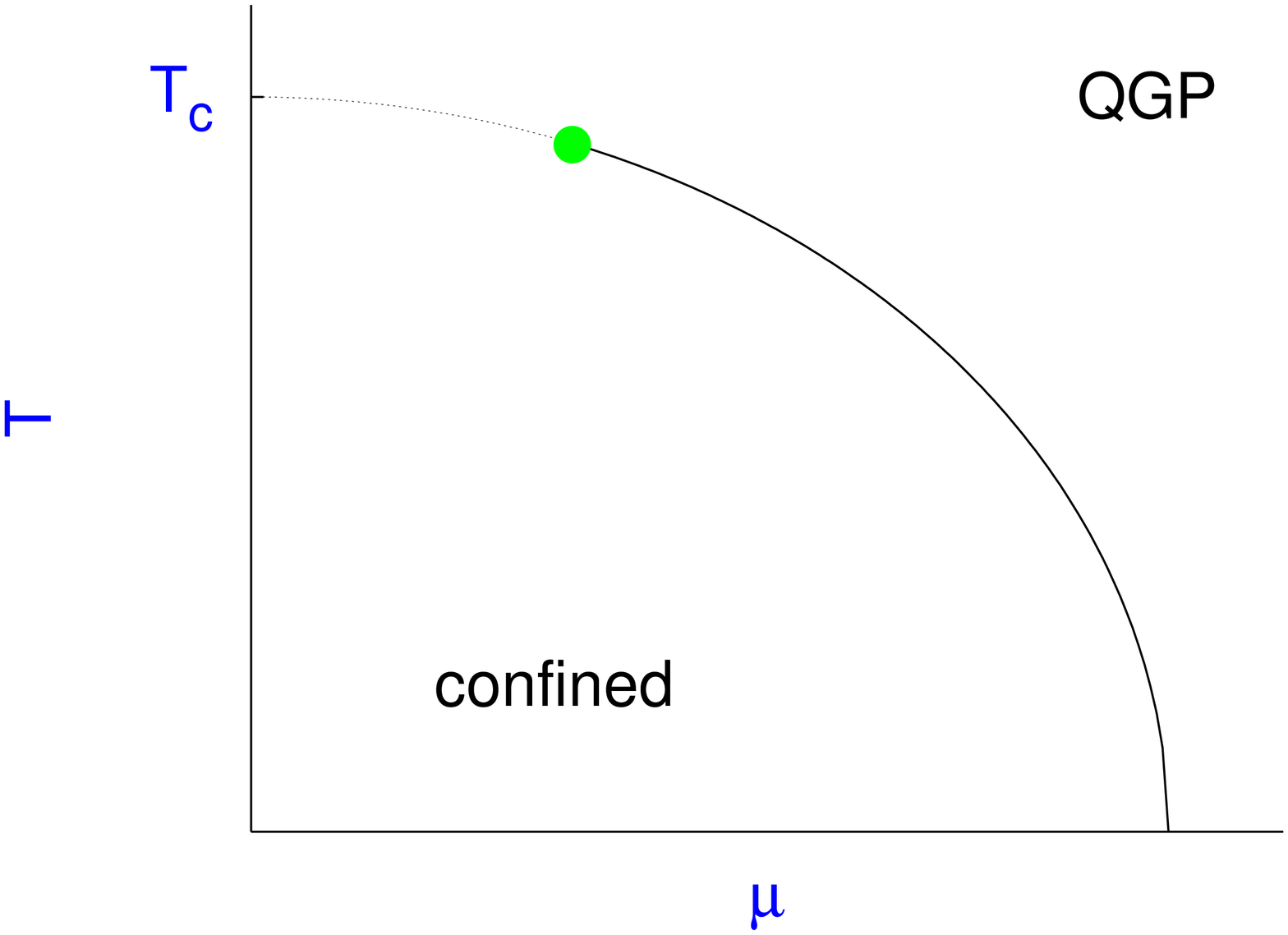} \\
\includegraphics[height=6.00cm,angle=-90]{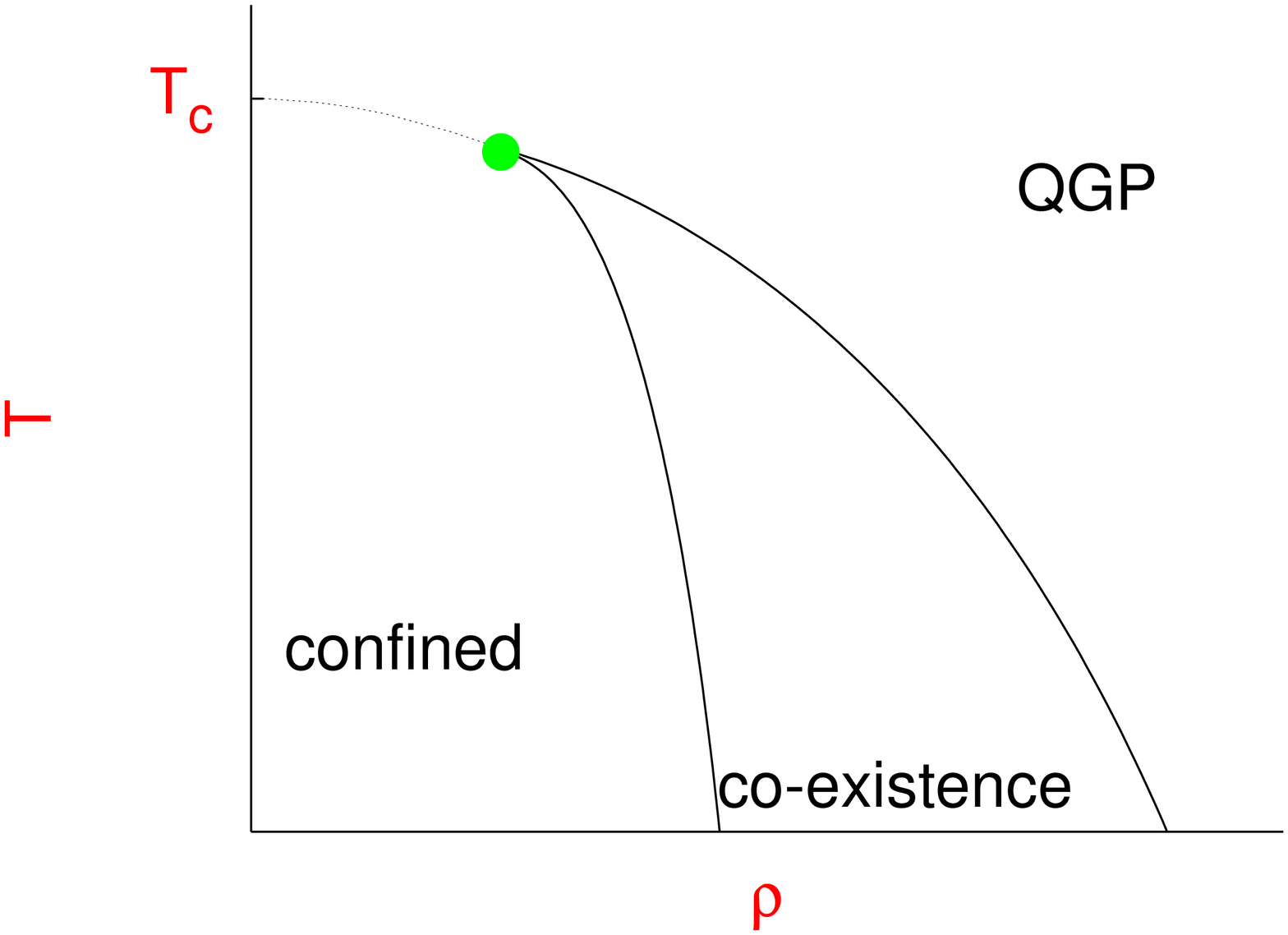}
\caption{Sketch of the conjectured QCD phase diagram in the grand-canonical 
({\em top}) and canonical ({\em bottom}) formalism.}\label{fig:simplifedQCD}
\end{center}
\vspace*{-0.8cm}
\end{figure}

The last few years have seen remarkable progress in the numerical study of QCD
at finite chemical potential $\mu$.
Still, the various methods~\cite{Fodor:2001au,Allton:2002zi,D'Elia:2002gd,Azcoiti:2005tv} suffer from
systematic uncertainties, which limit their range of reliability to about $\frac{\mu}{T} \lesssim 1.0$.
For a recent review, see Ref.~\cite{Philipsen}.
We try to address this apparent limitation by using a canonical approach~\cite{Kratochvila:2003rp,Alexandru:2005ix}, where
we focus on the matter density $\rho$, rather than the chemical potential. 
Fixing the baryon number brings us closer to the experimental situation
of a heavy-ion collision.
Moreover, our method is particularly appropriate to explore few-nucleon
systems at low temperature, and in principle, allows to study the bulk properties of nuclear matter and
the nuclear interactions. Here, we extend its use and determine the phase boundary between the confined phase and the quark
gluon plasma, as illustrated in Fig.\ref{fig:simplifedQCD}.

The top figure shows a sketch of the conjectured phase diagram of
QCD in the $T$-$\mu$ plane. At small chemical potential, the phase
transition is a rapid crossover, which ends in a second order
endpoint, followed by a first order transition. Correspondingly,
the bottom figure illustrates the transition in the $T$ - $\rho$
plane. The first order transition is manifested in a co-existence
region. In this proceedings, we describe how we identify the
co-existence region and how we determine the phase diagram in the
$T$-$\mu$ as well as in the $T$-$\rho$ plane.

\section{Partition Functions}
                                                                                
We construct the canonical partition function $Z_C(T,Q)$ by fixing the number ofquarks $\hat N=\int d^3 \vec x \; \bar{\psi}(\vec x) \; \gamma_0 \; \psi(\vec x)$
to $Q$. We insert a $\delta$-function in the grand canonical partition function
$Z_{GC}(T,\mu)$
\be
\int [DU][D\bar\Psi][D\Psi]\; e^{-S_g[U;T]-S_F[U,\bar \Psi, \Psi;T]} \delta\left(\hat N-Q\right) 
\ee  
\noindent
The $\delta$-function admits a Fourier representation $ \delta\left(\hat N-Q\right)=\int d\bar \mu_I \; e^{i \bar \mu_I (N-Q)}$.
We recognise $i\mu_I=i\bar \mu_I T$ as an imaginary chemical potential and exploit the $\frac{2 \pi T}{3}$-periodicity~\cite{Roberge:mm}
in $\mu_I$ of $Z_{GC}(T,\mu=i\mu_I)$
\bea
  Z_{C}(T,Q) = \frac{3}{2\pi} \int_{-\frac{\pi}{3}}^{\frac{\pi}{3}} d\bar \mu_I
 e^{-i Q \bar \mu_I }Z_{GC}(T,i\bar \mu_I T  ) \nonumber \\ 
\hspace*{0.5cm}
  \underset{Q = 3B}=  \frac{1}{2\pi} \int_{-\pi}^{\pi} d\left(\frac{\mu_I}{T} \right)  e^{-i 3 B \frac{\mu_I}{T} }Z_{GC}(T,i \mu_I) 
\eea
Thus, the canonical partition functions are the coefficients of the Fourier expansion
in imaginary $\mu$ of the grand canonical partition function.
As a consequence of the $\frac{2 \pi T}{3}$-periodicity, the canonical partition functions are zero for non-integer baryon number $B = Q/3$.
                                                                                
From the canonical partition functions, the grand canonical partition function can be reconstructed using the fugacity
expansion (in fact a Laplace transformation)
\bea \label{eq:laplace_trafo}
\hspace*{-0.5cm}
Z_{GC}(T,\mu)  \underset{V \to \infty}= \int_{-\infty}^{\infty} d \rho \;e^{3 V
\rho \frac{ \mu }{ T }} Z_C(T,\rho) \nonumber \\
   = \int_{-\infty}^{\infty} d \rho \;e^{-\frac{ V}{T} \left( f(T,\rho) -  3 \mu \rho  \right) }
\hspace*{0.2cm}
\eea
with the baryon density $\rho=\frac{B}{V}$
and the Helmholtz free energy density $f(T,\rho)=- \frac{T}{V} \log Z_C(T,\rho)$.
The relation between baryon density and chemical potential can be expressed as $\langle\rho\rangle(\mu)$
or $\mu(\rho)$:
\be
 \langle\rho\rangle(\mu) = \frac{1}{Z_{GC}(T,\mu)} \int d\rho\;\rho\;e^{3 V \rho \frac{ \mu }{ T }} Z_C(T,\rho) 
\ee
or
\be
    \mu(\rho) = \frac{1}{3}\frac{\partial f(\rho)}{\partial \rho} \;. \label{eq:B_mu}
\ee
While the first expression is exact in any volume, the second is obtained via a
saddle point approximation (exact in the thermodynamic limit)
and may have more than one solution when solving for the baryon density at a given chemical potential, see
Fig.\ref{fig:results_FB1B_maxwell_3_4.92}. We discuss this issue in detail below.
                                                                                
\section{Method}
                                                                                
Following \cite{Hasenfratz:ax}, we express the canonical partition
function in a ratio, which can be measured by Monte Carlo
simulation as an expectation value:
\bea
\frac{Z_C(B,\beta)}{Z_{GC}(\beta_0=\beta,\mu=i \mu_{I_0})} \hspace*{1.5cm} \nonumber \\
= \frac{1}{Z_{GC}(\beta_0,i \mu_{I_0})} \int [DU]\; e^{-S_g[U;\beta_0]} \det(U;i \mu_{I_0}) \nonumber \\
\times \frac{1}{2\pi} \int_{-\pi}^{\pi} d \left( \frac{ \mu_I }{ T }\right) \; e^{-i 3 B \frac{ \mu_I }{ T }}
\frac{ \det(U;i \mu_I) }{ \det(U;i \mu_{I_0}) } \nonumber \\
=
 \langle \frac{1}{2\pi} \int_{-\pi}^{\pi} d \left( \frac{ \mu_I }{ T }\right) \; e^{-i 3 B \frac{ \mu_I }{ T }}
\frac{ \det(U;i \mu_I) }{ \det(U;i \mu_{I_0}) } \rangle_{\beta_0, i\mu_{I_0}} \nonumber \\
\equiv\langle \frac{\hat Z_C(U;B)}{\det(U;i \mu_{I_0})}\rangle_{\beta_0, i\mu_{I_0}} \hspace*{1.5cm} \label{MCobs}
\eea
                                                                                
\noindent
where $Z_{GC}(\beta_0,i \mu_{I_0})$ is the grand canonical partition function sampled by ordinary Monte Carlo methods, here for notational simplicity at
$\beta_0=\beta$. The $\hat Z_C(U;B)$'s are the Fourier coefficients of the fermion determinant for a given configuration $\{U\}$.
Although the average in Eq.(\ref{MCobs}) should be real positive, the individual measurements are complex,
with a sometimes negative real part. This is how the sign problem manifests itself in our approach.
Moreover, a reliable estimate depends on a good overlap of our Monte Carlo ensemble with the canonical
sector $B$ at temperature $\beta$. We address this issue by following the idea of Ref.~\cite{Fodor:2001au}
and including both confined and deconfined configurations in our ensemble.
Indeed, we supplement the ensemble at $(\beta_c(\mu=0),\mu=0)$ with additional
critical ensembles at imaginary chemical potential, non-zero isospin chemical potential, and ensembles generated with an asymmetric
Dirac coupling~\cite{Azcoiti:2005tv}
- in principle, any ensemble is allowed.
We then combine all this information about a particular canonical partition function $Z_C(B,\beta)$ by Ferrenberg-Swendsen
reweighting~\cite{Ferrenberg:yz}.
                                                                                
The Fourier-coefficients of the determinant $\hat Z_C(U;Q)$ are calculated exactly~\cite{Hasenfratz:ax}.
In the temporal gauge ($U_4({\mathbf x},t)={\mathbf 1}$ except for $t=N_t-1$), the staggered fermion matrix $M$ in the
presence of a chemical potential can be written in the form
\small
\begin{eqnarray*}
  \left( \begin{matrix}
        B_0 & {\mathbf 1} & 0 & ... & 0 & U^\dagger_{N_t-1} e^{-\mu a N_t}\\
         - {\mathbf 1} & B_1 &  {\mathbf  1} & 0 & ... & 0\\
         0 & - {\mathbf  1} & B_2 &  {\mathbf  1} &  0& ... \\
          &  &  &   ... &  & \\
         -U_{N_t-1}e^{\mu a N_t} & 0 & ... & 0 &  - {\mathbf  1} &  B_{N_t-1}
      \end{matrix}
  \right) 
\end{eqnarray*}
\begin{align}
\longleftrightarrow ~~~~ &P = \left(
        \prod_{j=0}^{N_t-1} \left( \begin{matrix}
            B_j & {\mathbf  1} \\
            {\mathbf 1} & 0
            \end{matrix} \right)
       \right) U_{N_t-1} \notag\;,
\end{align}
\normalsize
where the $B_i$'s contain all space-like contributions. Ref.~\cite{Gibbs:1986hi} showed that the determinant can be
computed for any chemical potential at the cost of diagonalising the so-called ``reduced matrix''
$P$. The determinant is given in terms of $P$'s eigenvalues $\lambda_1, \ldots,
\lambda_{6V}$, where $V$ is the spatial volume:
\bea
  \det M(U;\mu) =  e^{3 V\;\mu a N_t}\; \prod_{i=1}^{6 V}\left( \lambda_i + e^{- \mu a N_t} \right) \nonumber \\
   = \sum_{Q=-3 V}^{Q=3 V} \hat Z_C(U;Q) e^{-Q \mu a N_t} \;.
\eea
Matching term by term, we then solve for the Fourier coefficients $\hat Z_C(U;Q)$. This delicate step requires a special multi-precision library.
The diagonalisation of the reduced matrix $P$ is computationally intensive and takes ${\cal O}(V^3)$ operations.

\section{Results} 
 
We study the Helmholtz free energy $F(B)\equiv -T \log \frac{Z_C(B)}{Z_C(0)}$ in a theory of
four degenerate flavors of staggered quarks with mass $m a = 0.05$ ($\frac{m}{T}=0.2$, $m_\pi \sim 350$ MeV) on a
small, coarse $6^3 \times 4$ lattice with volume $\sim (1.8$fm$)^3$. For the quark mass
we chose, the phase transition is first order
at $\mu=0$, and presumably remains first order for all chemical potentials.
 
\begin{figure}[!hht]
\hspace*{1.5cm} 
\includegraphics[height=5.00cm,angle=-90]{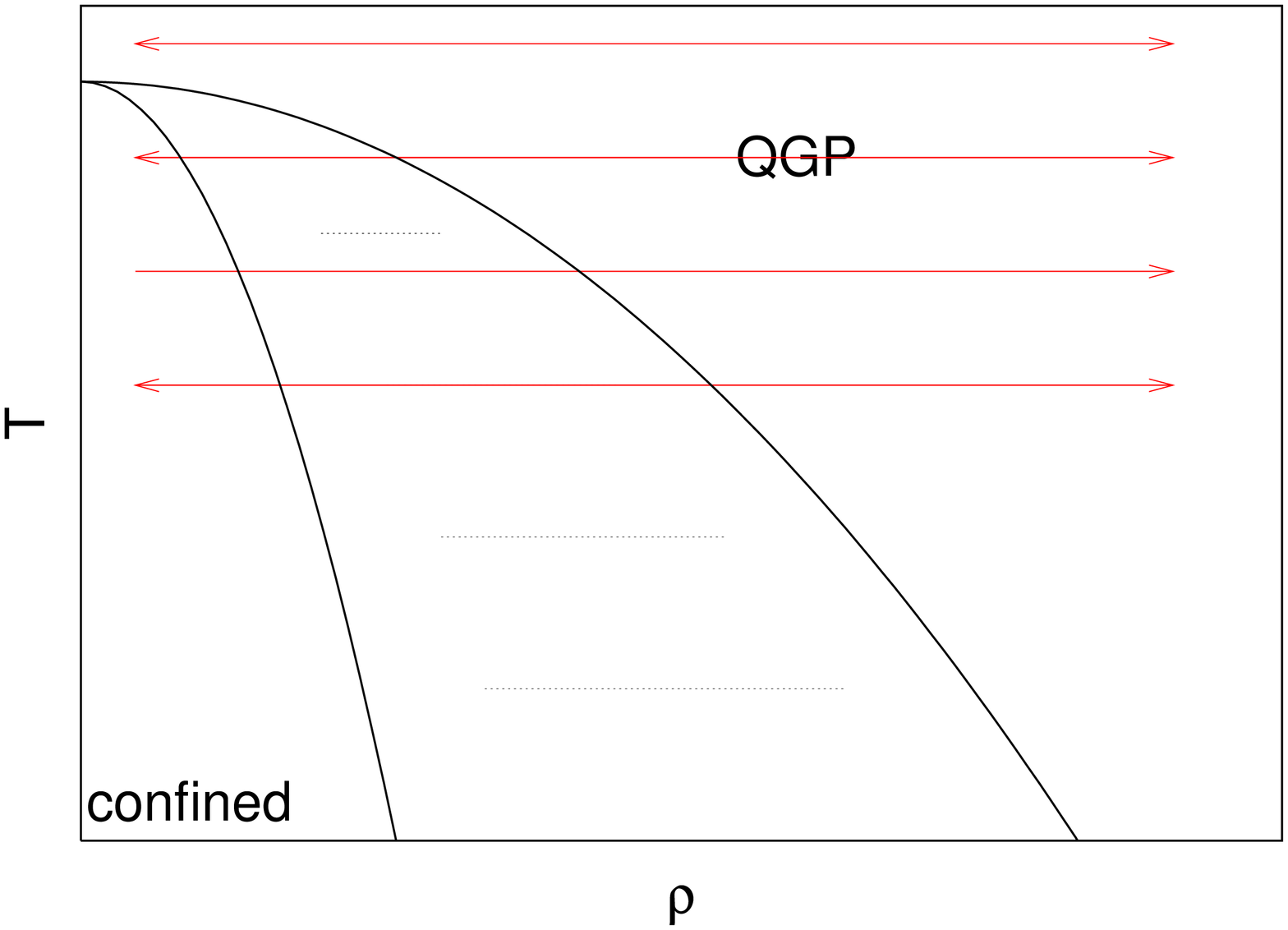} \\
\hspace*{-0.3cm} 
\includegraphics[height=8.00cm,angle=-90]{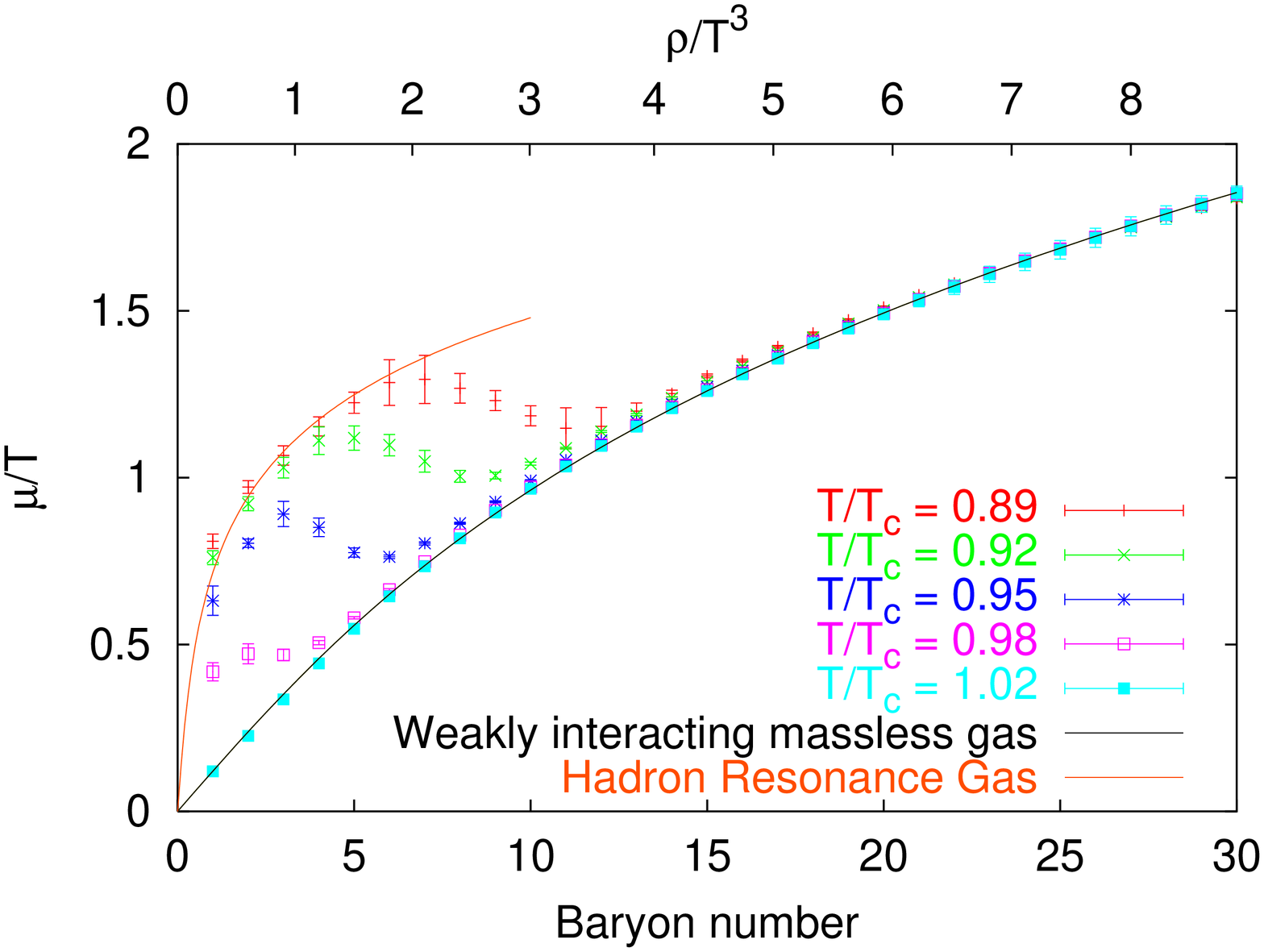}
\caption{({\em top}) A sketch of the ``scans'' in our phase diagram. ({\em bottom}) The derivative of the free energy at fixed temperature as a function
of the baryon number (baryon density). In the saddle point approximation, the $y$-axis is $\frac{\mu}{T}=\frac{F(B)-F(B-1)}{3T}$.}
\label{fig:results_FB1B_B}
\end{figure}

\begin{figure}[!htb]
\begin{center}
\includegraphics[height=7.00cm,angle=-90]{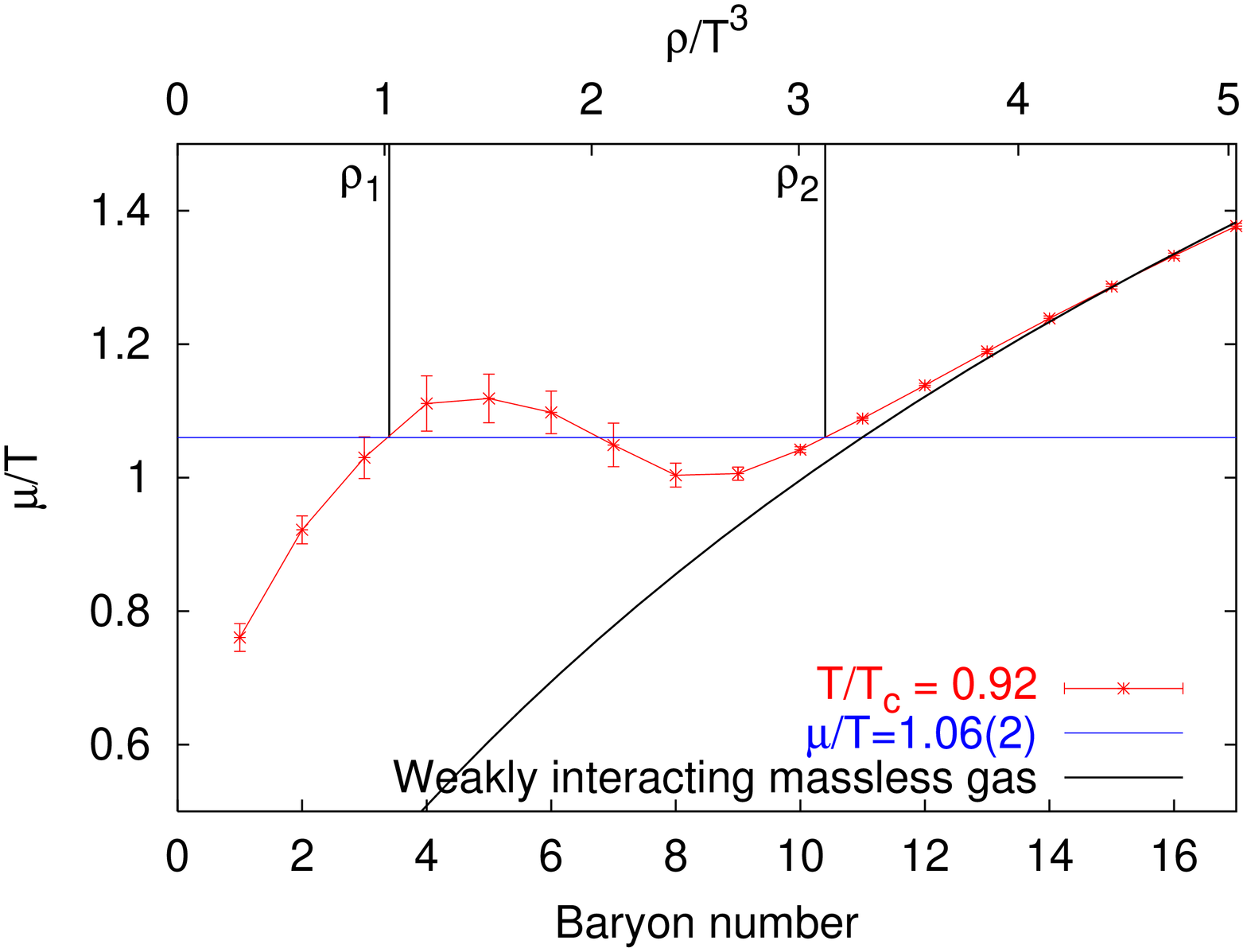}
\includegraphics[height=7.00cm,angle=-90]{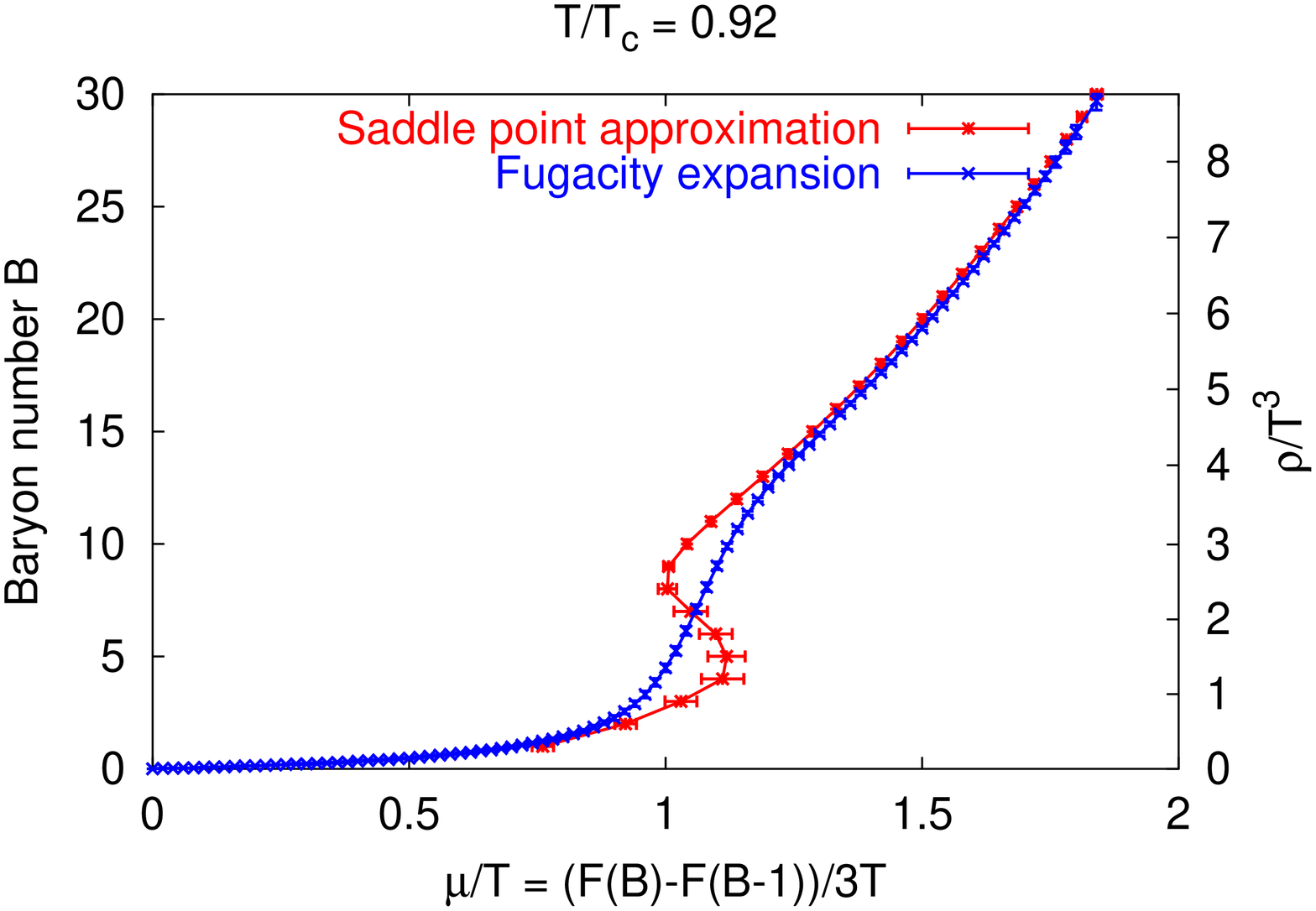}
\caption{({\em top}) The Maxwell construction allows to extract the critical chemical potential and the boundaries of
the co-existence region. ({\em bottom}) Comparing the saddle point approximation (red)
with the fugacity expansion (blue).
Strong finite-size effects in the latter obscure the first-order transition.}
\label{fig:results_FB1B_maxwell_3_4.92}
\end{center}
\end{figure}

We  ``scan'' the phase diagram by varying the baryon density at fixed temperature, see Fig.\ref{fig:results_FB1B_B}.
We measure $\frac{F(B)-F(B-1)}{3T}$ and assume the validity of the saddle point
approximation to
equate this quantity with $\frac{\mu(B)}{T}$ following Eq.~(\ref{eq:B_mu}). This assumption will be tested in Fig.\ref{fig:results_FB1B_maxwell_3_4.92}. 
 
Note that accurate results are obtained, up to high densities ($> 5$ baryons/fm$^3$) and large chemical
potentials ($\frac{\mu}{T} \sim 2$).
The first order phase transition and the associated metastabilities are clearly
visible in the ``S-shape'' of $\frac{\mu}{T}(B)$. The low-density regime can be
reasonably well described by a simple hadron resonance gas
Ansatz, $\frac{\rho}{T^3}=3 f(T) \sinh( 3\frac{\mu}{T} )$ with $f(T)$ as the only free parameter. The high-density regime
almost corresponds to a gas of free massless quarks $\frac{\rho}{T^3} = N_f \left(\frac{\mu}{T} \right) + \frac{N_f}{\pi^2} \left(\frac{\mu}{T} \right)^3$
when taking cut-off corrections~\cite{Allton:2003vx} into account. The solid line in Fig.\ref{fig:results_FB1B_B}(bottom)
is obtained by fitting the linear and cubic terms in this expression. Instead of the free value 1, the fitted coefficients are
$0.82(2)$ and $1.94(6)$ respectively. Thus, the equation of state for the quarks in the plasma phase
differs little from the Stefan-Boltzmann law. This has been observed also in \cite{Csikor:2004ik} and \cite{Allton:2005gk}. 
Such a result is surprising, given the experimental evidence for a strongly
interacting plasma. Perhaps it simply reflects the lack of sensitivity of
the quark equation of state to the coupling strength. Note in particular 
the striking independence of $\frac{\rho}{T^3}(\frac{\mu}{T})$ 
on temperature, as long as one is in the plasma phase. 
                                                                                
For a given temperature $T$, we identify the boundaries $\rho_1$ and $\rho_2$ of the co-existence region and the critical chemical potential $\mu_c$
as follows.
Equality of the free energy densities in the two phases, $f(\rho_1)-3 \mu_c \rho_1 = f(\rho_2)-3 \mu_c  \rho_2$,
implies
\be \label{eq:max_critical}
 \int_{\rho_1}^{\rho_2} d\rho (f'(\rho)-3 \mu_c) = 0\;.
\ee  
\noindent
Since $f'(\rho)$ is the quantity measured in Fig.\ref{fig:results_FB1B_B}, we determine $\rho_1, \rho_2$ and $\mu_c$
by a ``Maxwell construction'' illustrated in Fig.\ref{fig:results_FB1B_maxwell_3_4.92} (top)
for the temperature $\frac{T}{T_c}=0.92$. The value of $\frac{\mu}{T}$ defining
the horizontal line is adjusted
to make the areas of the two ``bumps'' in the S-shape equal.
The two outermost crossing points define $\rho_1$
and $\rho_2$, the boundaries of the co-existence region. Here, $\frac{\mu_c}{T}=1.06(2)$ is the value of the critical
chemical potential. 
                                                                                
We can cross-check this result by making use of the fugacity
expansion Eq.(\ref{eq:B_mu}), see Fig.\ref{fig:results_FB1B_maxwell_3_4.92} (bottom). For a given chemical
potential, we measure the baryon number $\langle B\rangle(\mu)$. We see a 
rapid variation
at the same value $\frac{\mu_c}{T} \approx 1.06$, but the rounding due to finite size effects is very strong.
In contrast, our criterion for criticality (equality of the free energies)
gives a clear first-order signal.
In the thermodynamic limit, both approaches will indicate the same jump at 
$\mu_c$.

\begin{figure}[!htb]
\begin{center}
\includegraphics[width=7.00cm,angle=0]{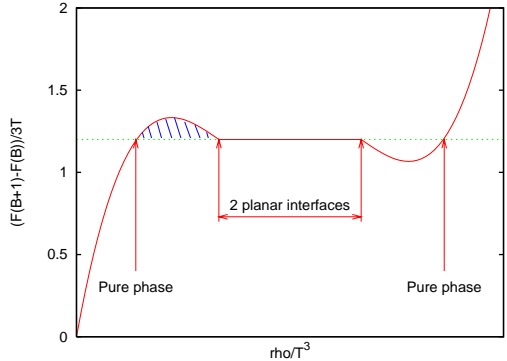}
\caption{Connection between the S-shape of the derivative of the free energy
and the interface tension: the shaded area represents the free energy required
to create two planar interfaces.}
\label{fig:tension}
\end{center}
\end{figure}

Thus, the S-shape we observe is a finite-size effect. Its origin is clarified
in Fig.\ref{fig:tension}. In a system of spatial size $L$, it takes a free
energy $2 L^2 \sigma/T$ to create two planar, static interfaces separating 
the two phases, where $\sigma$ is the interface tension. Once these two
non-interacting interfaces are present, the baryon density can be varied
at constant free energy by changing their separation. Thus, the free energy
required for the creation of these interfaces is equal to the shaded area
in Fig.\ref{fig:tension}. A crude trapezoidal integration gives a reasonable
value for the reduced interface tension $\sqrt{\sigma/T} \sim 35-45$ MeV.
The contribution of these interfaces to the free energy density goes as
$1/L$, and vanishes in the thermodynamic limit.

\begin{figure}[!thb]
\begin{center}
\hspace*{0.0cm}
\includegraphics[height=7.70cm,angle=-90]{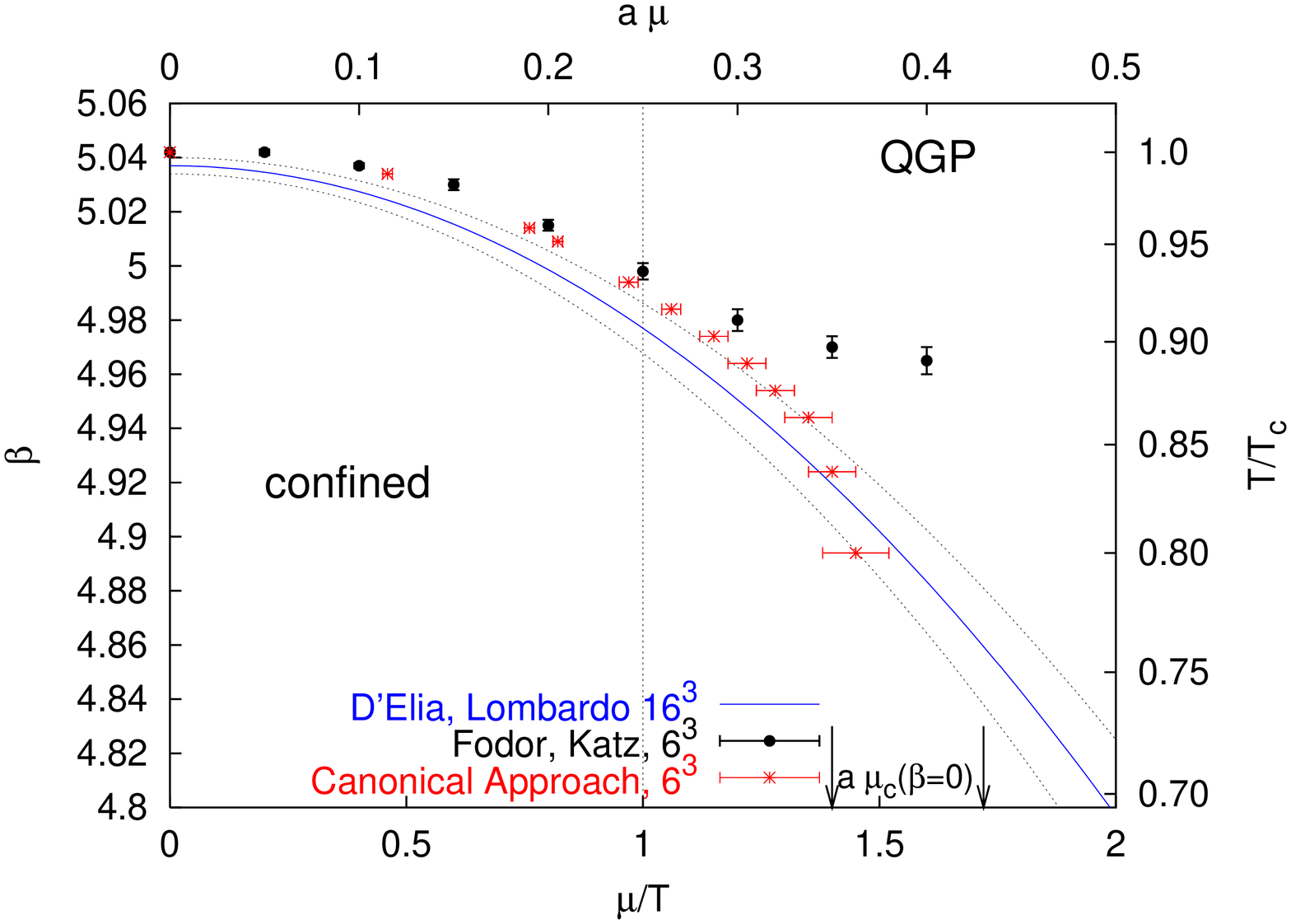} \\
\hspace*{-0.5cm}
\includegraphics[height=7.50cm,angle=-90]{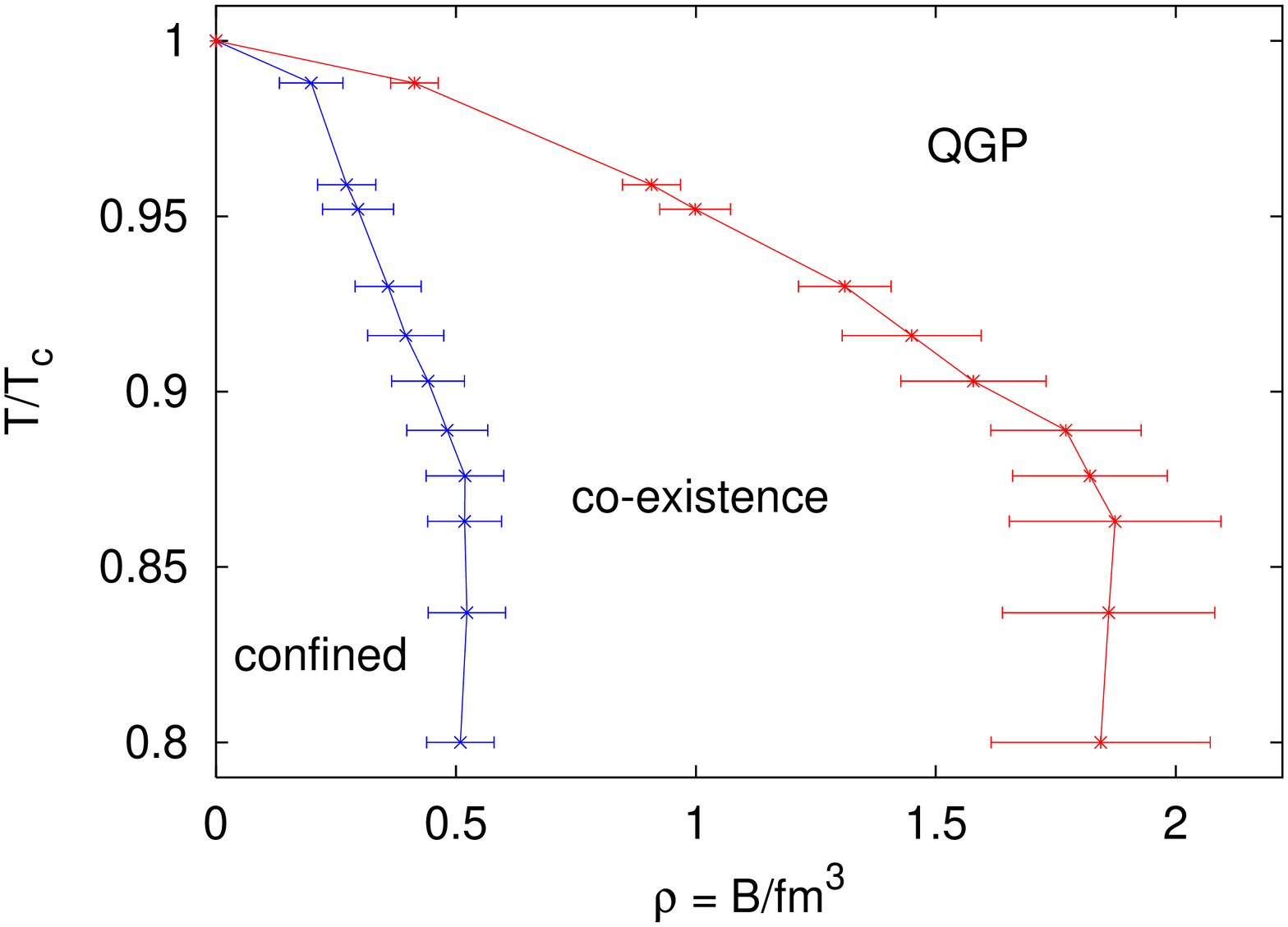}
\caption{({\em top}) The phase diagram in the $T$-$\mu$-plane, obtained by
3 different methods. The two arrows indicate the range of the strong coupling
($\beta=0$) prediction~\cite{Kawamoto}. ({\em bottom}) The phase diagram 
which we obtain, in the $T$-$\rho$-plane.}
\label{fig:phasediags} 
\end{center}
\end{figure}

In Fig.\ref{fig:phasediags} we present the phase diagrams in the $T$-$\mu$ as well as in the $T$-$\rho$ plane.
The top figure summarizes results from various methods, all for the same theory: 4 flavors of staggered quarks with $a m = 0.05$, $N_t=4$ time-slices;
only the spatial volume varies as indicated.
All methods agree nicely up to $\mu/T \sim 1$, but have different systematics
which become dominant at larger chemical potentials.
The multi-parameter approach of \cite{Fodor:2001au} uses a single Monte Carlo
ensemble generated at $(\beta_c,\mu=0)$, and determines the phase transition
from Lee-Yang zeroes. To understand the systematic errors involved, we have
repeated this study, but used instead the peak of the specific heat to
determine the phase transition. Our results are completely consistent with
\cite{Fodor:2001au}. However, we have noticed that the ``sign problem'' grows 
dramatically more severe with increasing chemical potential, but that
our statistical error, based on jackknife bins as in \cite{Fodor:2001au},
fails to reveal this breakdown of the method.
                                                                                
The parabolic fit~\cite{D'Elia:2002gd} is consistent with the more elaborate
approach of \cite{Azcoiti:2005tv}. Both methods perform an
analytic continuation from imaginary $\mu$, for which the systematic errors are
hard to quantify. Our new results are shown in red. There is no
strong inconsistency with other results, but we observe a clear sign of bending
down starting at $\frac{\mu}{T}\sim 1.3$.
In fact this  must happen, if the critical line is to fall within the range
$a \mu_c=0.35..0.43$ at $\beta=0$, as predicted from a strong coupling
analysis~\cite{Kawamoto}.

Fig.\ref{fig:phasediags}(bottom) shows the first phase diagram in 
the $T$-$\rho$ plane. The densities at the boundaries of the co-existence region seem to remain constant for $T \lesssim 0.85 T_c$ already,
with $\rho_{\textnormal{QGP}} = 1.8(3) B/$fm$^3$ and $\rho_{\textnormal{confined}}= 0.50(5) B/$fm$^3$.
The latter is a plausible value for the maximum nuclear density in our 4-flavor, $m_\pi=350$ MeV QCD theory.

\section{Conclusions} 
                                                                                
We study QCD in a canonical framework,
which is promising for the study of few-nucleon systems at low temperature, but
proves also capable of exploring high density regimes ($\mu/T \lesssim 2$) at temperatures $T \gtrsim 0.8 T_c$.
We have determined the phase boundary between
the confined phase and the quark gluon plasma in both the $T$-$\rho$ and the $T$-$\mu$ plane.
In the latter, our results are in agreement with the literature, however we observe a bending down
of the critical line at $\frac{\mu}{T} \sim 1.3$.
The two phases can be rather well described
by the hadron resonance gas at low densities and by a weakly interacting massless gas
at high densities.
                                                                                
\noindent
\section{Acknowledgements}

Ph.~de~F. thanks the organizers for a smoothly run, stimulating conference.
We thank the Minnesota Supercomputer Institute for computing resources.

\end{document}